# Cepstral Strain Mapping for Small Pixel-Count Detectors

Harikrishnan KP[1,*], Dasol Yoon[1,2], Yu-Tsun Shao[1,3], Zhaslan Baraissov[1], Luigi Mele[4], Christoph Mitterbauer[4], Erik Kieft[4], Stefano Vespucci[4], David A. Muller[1,5,*]

[1]School of Applied and Engineering Physics, Cornell University, Ithaca, NY, USA
[2]Department of Materials Science and Engineering, Cornell University, Ithaca, NY, USA
[3]Mork Family Department of Chemical Engineering and Materials Science, University of Southern California, Los Angeles, CA, USA
[4]Thermo Fisher Scientific, Eindhoven, Netherlands
[5]Kavli Institute at Cornell for Nanoscale Science, Cornell University, Ithaca, NY, USA
*Corresponding Authors: i) David A. Muller, E-mail: david.a.muller@cornell.edu; ii) Harikrishnan KP, Email: hk944@cornell.edu

## Abstract

With the decreasing sizes of integrated-circuit components, the semiconductor industry is in growing need of high-throughput strain mapping techniques that offer high precision and spatial resolution, with desired industry goals of 0.01-0.1% and 1 nm respectively. As the fundamental limitation on the measurement precision is set by the Poisson noise, pixel array detectors with high saturation current, high dynamic range and fast readout are ideally suited for this purpose. However, due to the limited pixel count on these detectors, they do not work well with traditional strain mapping algorithms that were optimized to work on datasets with a large pixel count. Here, we evaluate the cepstral transform that was designed to address this problem, with the precision determined by the convergence, collection angles and dose while remaining insensitive to the pixel count. We test the performance of our method on silicon wedges and Si-SiGe multilayers, and using datasets collected at different conditions, we show how the measured strain precision scales as a function of dose, aperture size and sample thickness. Using precession gives a further improvement in precision by about 1.5-2x, whereas energy filtering does not have a significant impact on the cepstral method for device-relevant sample thickness ranges.

**Introduction**

Pixel array detectors like the EMPADs (Tate et al., 2016; Philipp et al., 2022) and DECTRIS ARINA (Zambon et al., 2023; Stroppa et al., 2023) are gaining widespread adoption for high speed and high current diffraction measurements required for ptychography (Nellist et al., 1995; Jiang et al., 2018; Chen et al., 2021) and other 4D-STEM techniques (Ophus, 2019). However, their limited pixel count (128 x 128 for EMPADs and 192 x 192 or 96 x 96 for ARINA) presents a challenge for strain mapping applications, as common strain mapping algorithms like template cross-correlation (Müller et al., 2012; Pekin et al., 2017; Zeltmann et al., 2020; Savitzky et al., 2021) and edge-detection (Müller et al., 2012; Yuan et al., 2019) are designed to work on diffraction data collected with a large pixel-count detector. However, precise tracking of diffraction disk shifts for strain mapping does not fundamentally depend on a large pixel count. Numerical simulations of center-of-mass calculations have shown that the precision of disk detection improves with increasing pixel count only up to a disk diameter of 5–6 pixels, with diminishing returns beyond (Han et al., 2018; Nguyen et al., 2022). Therefore, the use of pixel array detectors for strain mapping requires the design of algorithms that can effectively utilize their high dynamic range, fast readout speed and large saturation current while working efficiently even with a low pixel count. The Exit Wave Power Cepstral (EWPC) transform (Padgett et al., 2020) was developed to meet these demands, with its performance determined by the maximum dose, and the largest usable angle, rather than the number of detector pixels. In addition to strain mapping applications (Padgett et al., 2023; Kavle et al., 2024), the cepstral transform has also been used for the imaging of lattice distortions using the diffuse scattering signal in the diffraction pattern (Shao et al., 2021), classification of microstructure variations (Zhang et al., 2020; Yoo et al., 2024; Sun et al., 2023; Kp et al., 2025;

Ghanbari et al., 2025), and identifying short-range order in high-entropy alloys, where it serves as a useful pre-processing step for subsequent data mining (Hsiao et al., 2022; Cueva et al., 2018). In this study, we focus on evaluating how EWPC performs under different experimental conditions and exploring its applicability for high-throughput strain mapping applications.

The cepstral transform has its origins in audio signal processing (Bogert B. P. et al., 1963) where it is widely used for deconvolving a slowly-varying envelope function from the carrier wave and has broad applications in autotuning and speech analysis. A direct cepstral transform would start with what in electron microscopy would be the ideal real-space exit wave and then calculating the power spectrum, a step that is accomplished in EWPC by forming the diffraction pattern[16]. In other words, the cepstral transform here is on the exit wave, not the diffraction pattern. Although the experiment begins with the collection of the diffraction pattern, this distinction motivates the name EWPC rather than simply 'cepstral transform,' even though it is often colloquially referred to as cepstral STEM. When EWPC is adapted for strain mapping applications, the lattice plays the role of the carrier wave, whereas variations in sample thickness and tilt effectively form an envelope function attenuating the carrier wave, provided they vary on a length scale much larger than the lattice spacing (Padgett et al., 2020). Hence, this transformation helps decouple the influence of thickness and tilt variations from changes in the lattice spacing, thereby reducing diffraction artifacts in strain maps. In contrast to typical methods that rely on precisely tracking the diffraction disks, the cepstral transform leverages the periodicity in the diffraction pattern making it effective even with only a few pixels per diffracted disk as in pixel array detectors, also leading to more compact, memory-efficient data sets. As processing times scale with the number of detector pixels, the ability to use small datasets considerably speeds up the computing time. Moreover, this technique does not require any pre-processing and does not have software parameters that need to be tuned according to the experimental conditions under which the data is collected. These factors along with the ease of parallelization make the combination of EWPC technique and pixel array detectors potentially well suited for high-throughput applications and live processing of strain maps during data acquisition.

Here we analyze factors that set limits on the achievable precision in strain mapping and identify two noise regimes – i) a Poisson-noise limited regime at low dose and ii) a dynamical-diffraction limited regime at high dose. In the former, precision is fundamentally limited by the Poisson/shot noise and leads to a $\frac{\alpha}{\sqrt{N}}$ scaling, where $\alpha$ is the numerical aperture or convergence semi-angle and N is the number of electrons, i.e. the dose used in each measurement. The linear scaling with $\alpha$ represents the trade-off between precision and spatial resolution ($\lambda/\alpha$) while the $\frac{1}{\sqrt{N}}$ scaling highlights the need for detectors capable of collecting the largest possible dose in the shortest possible time. From this perspective, pixel array detectors are well-suited for strain mapping because of their large saturation currents and high frame rates while maintaining single electron sensitivity and high detective quantum efficiency (DQE). In contrast, scintillator-based detectors have a low DQE from the electron-photon conversion and record noisy diffraction patterns, while monolithic active pixel sensor (MAPS) detectors maintain high DQE only when operating at low count rates that in turn limits the total dose (Li et al., 2013; McMullan et al., 2014). This improvement in precision with increasing dose plateaus after a critical value, beyond which dynamical diffraction contrast within the diffraction disks limits both precision and accuracy. We demonstrate how further improvements in precision and accuracy can be achieved by using precession electron diffraction (Vincent & Midgley, 1994; Midgley & Eggeman, 2015; Cooper et al., 2015; Rauch et al., 2010; MacLaren et al., 2020) that can reduce dynamical diffraction artifacts. Our preliminary strain mapping results on silicon wedge samples with EWPC were reported in conference proceedings (Harikrishnan et al., 2021; Yoon et al., 2022). We extend these here, including to a SiGe multilayer.

**Materials and Methods**

*Experiments and Simulation*

The samples used in the study were prepared using a Thermo Fisher Helios G4 UX focused ion beam, with initial milling at 30 kV followed by final thinning at 2-5 kV. The 4D STEM datasets used in the study were acquired using the EMPAD or EMPAD-G2 detector installed on a Thermo Fisher Spectra 300 X-CFEG STEM or FEI Titan Themis (S)TEM at 300 kV.

The multislice simulations in this work were done using the MuSTEM simulation suite (Allen et al., 2015). The simulations were done for an unstrained silicon crystal oriented along the [110] zone axis and at an accelerating voltage of 300 kV with 8 slices per unit cell along the beam direction. Thermal diffuse scattering was modelled using quantum excitations of phonons (QEP) method (Forbes et al., 2010) in MuSTEM. The Bloch wave simulations were done using the EMAPS package (Zuo & Mabon, 2004).

*Strain mapping using the EWPC transform*

The exit wave power cepstrum (EWPC) is a non-linear transformation applied on diffraction patterns that results in a real-space representation of the distances in the crystal structure being probed (Padgett et al., 2020). It attempts to separate the lattice spacings from a slowly-varying envelope function that contains the tilt and thickness variations, thus reducing sensitivity to diffraction artifacts - something we test below using a wedge-shaped sample. The EWPC pattern is obtained by taking the Fourier transform of the logarithm of the diffraction pattern. It results in a pair correlation function similar to the Patterson function (Patterson, 1934) (with some important differences discussed in the appendix of Padgett et al., 2020) with peaks corresponding to projected inter-atomic distances of the crystal lattice structure as illustrated in Fig. 1 for simulated diffraction patterns of 30 nm thick Si [110] crystal with a semi-convergence angle of 1.2 mrad. The logarithm of the simulated position-averaged convergent beam electron diffraction (PACBED) pattern is shown in Fig. 1(a). The EWPC pattern generated from this diffraction pattern is given in Fig. 1(b) with the projected inter-atomic distances marked for comparison with the model of the Si [110] lattice in Fig. 1(c). The yellow, red and green arrows correspond to projected inter-atomic distances of 3.33 Å, 3.84 Å and 5.43 Å, respectively. Although not labelled, the first peak along the green line corresponds to the 1.36 Å dumbbell spacing of the (004) planes.

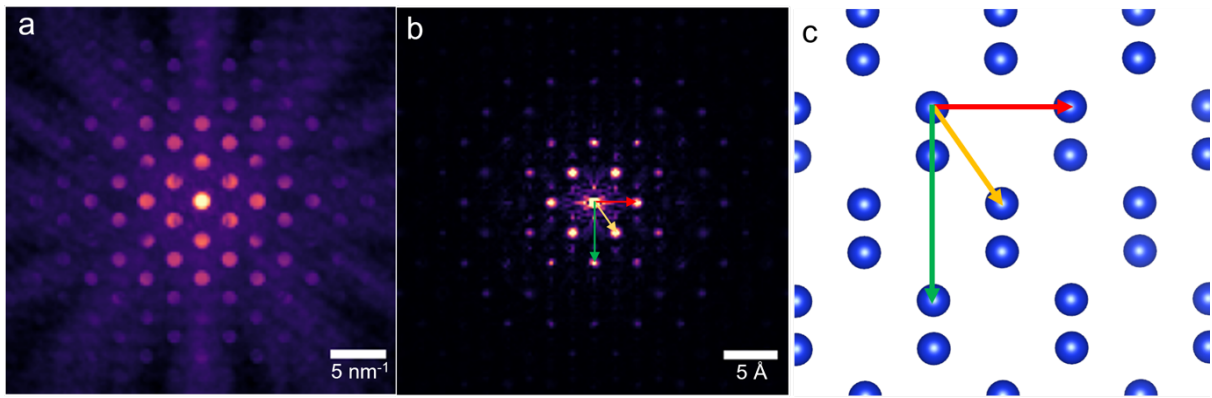

**Fig. 1. Illustration of EWPC transform as a pair-correlation function.** (a) shows the logarithm of simulated PACBED pattern for 30 nm Si[110] with 1.2 mrad semi-convergence angle at 300 kV. The corresponding EWPC pattern is given in (b) with the arrows denoting different projected inter-atomic distances in the lattice structure as shown in (c). The projected distances are 3.33 Å (yellow), 3.84 Å (red) and 5.43 Å (green). Although not labelled, the first peak along the green arrow in (b) corresponds to the dumbbell spacing in (c).

Strain mapping on a 4D-STEM dataset with the EWPC pattern is done by tracking the peak positions of a pair of EWPC spots for every diffraction pattern with sub-pixel precision. As mentioned earlier, the EWPC pattern is calculated as the discrete Fourier transform (using FFT algorithms) of the logarithm of the CBED pattern. As such, precision would be limited by the number of pixels in the FFT. However, higher precision can be obtained by exploiting interpolation properties of the Nyquist sampling theorem for a bandwidth-limited function and defining a Fourier transform function that acts on the discretely sampled CBED pattern to calculate the Fourier integral at all intermediate cepstral space points (instead of only the discretely sampled points), giving a continuous function in cepstral domain. Hence this function can be passed to an optimization algorithm to calculate positions of maxima or minima in cepstral space with sub-pixel accuracy. In our case, we are interested in finding the peak positions for a pair of EWPC spots (a brief discussion on peak-finding methods is provided in Supplementary Text 1), and hence a small region surrounding the EWPC spots in the discrete EWPC pattern is passed to the optimization algorithm to find the position of maximum intensity. The size of the region is manually chosen here (can also be automated using a peak-finding and thresholding algorithm if needed) to include variations in the position of the spot of interest but not too big to include the effects of the tails from other spots.

Suppose the peak positions for the 2 EWPC spots calculated at a real-space pixel (x,y) is given by the 2x2 matrix **P (x,y)**. Here, the 2 columns of the matrix correspond to the position in pixel units of the peak positions for the 2 EWPC spots. By comparing the matrix **P(x,y)** to a reference matrix **P$_0$** (which represents the reference of zero distortion) one can extract information about the relative distortions at each real space position. This information is encoded in the distortion matrix **D(x,y)**, which represents the matrix transformation between the point pairs given by **P(x,y)** and **P$_0$**. The distortion matrix includes the effect of both strain as well as rotation; these can be decoupled using a polar decomposition $\mathbf{D(x,y)} = \mathbf{R(x,y) \cdot U(x,y)}$. The in-plane rotation of the lattice θ is obtained from $\mathbf{R}\left(\theta = \tan^{-1}\frac{R_{21}}{R_{11}}\right)$, whereas the strain components $\varepsilon_{ij}$ can be obtained from $\boldsymbol{\varepsilon} = \mathbf{U} - \mathbf{I}$, where **I** is the 2x2 identity matrix. If the values of the projected atomic distances are known, they can be directly used to determine the reference matrix **P$_0$**. However, such an absolute reference could lead to complications in the analysis of experimental data, as the pixel calibrations may not be accurate, and the patterns could be at arbitrary rotations. Using the Lagrange strain approach (Zuo & Spence, 2016) to set a relative reference derived from the data itself offers a simple way around this problem.

*Optimal camera length*

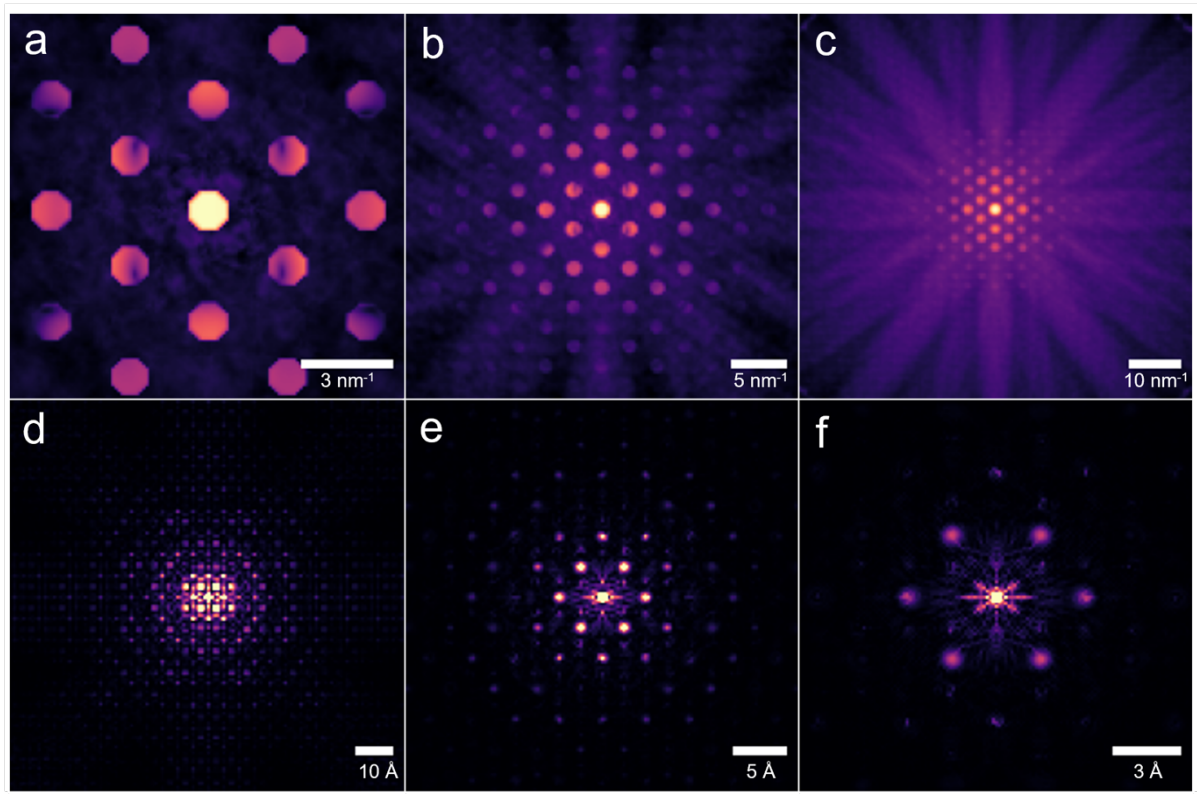

**Fig. 2. Comparison of EWPC patterns at different camera lengths**. The top panel (a,b,c) shows the logarithm of simulated PACBED patterns for 30 nm thick Si[110] for 3 different camera lengths with 1.2 mrad semi-convergence angle at 300 kV. The corresponding EWPC patterns (d,e,f) are displayed in the bottom panel. The pixel value range in the EWPC patterns have been clipped for better visualisation. The maximum collection angles along the x and y axes corresponding to these simulated patterns would roughly be (a,d)13 mrad, (b,e)35 mrad, and (c,f)75 mrad.

As discussed earlier, the peaks in the EWPC pattern correspond to directions of strong periodicity in the diffraction pattern. Therefore, to obtain the optimal EWPC patterns, the goal would be to set up the camera length to collect as many diffraction spots as possible on the detector without wasting any detector area beyond the Zeroth Order Laue Zone (ZOLZ).

Precession can help to increase the accessible ZOLZ, as can operating at a higher beam voltage where the Ewald sphere is flatter. Figure 2 shows CBED patterns simulated for a 30 nm thick Si[110] crystal (semi-convergence angle of 1.2 mrad) at 3 different camera lengths in the upper panel and the corresponding EWPC patterns in the lower panel. In Fig. 2(a), only a few ZOLZ spots are collected on the detector, and each EWPC spot is derived from the periodicity of only a few Bragg spots. The sampling in cepstral space is lower than desired, and the EWPC spots of interest are very close to the bright central spot (DC component). The other extreme case is shown in Fig. 2(c), where all the ZOLZ spots are collected in a small area near the center of the detector and a large area of the detector goes unused and increasing the risk of saturating diffraction peaks. The optimum case would be Fig. 2(b), where all the ZOLZ spots are collected without wasting any detector pixels. The EWPC spots corresponding to the projected inter atomic distances of interest are now well resolved and separated from the central spot.

*Scaling of Precision with Numerical Aperture and Dose*

The standard deviation in the measurement of the shift of a diffracted disk or equivalently an EWPC spot for a simple two-pixel model can be derived analytically. Chapman et al. (Chapman et al., 1978) has shown that the minimum deflection angle $\theta$ of a constant intensity disk that can be detected on a two-pixel detector is $\theta = \frac{\pi}{4}\frac{\alpha}{\sqrt{N}}$ where $\alpha$ is the total angular spread of the disk (convergence angle) and N is the total dose. Hence, the precision of strain measurement would scale as:

$$\delta\varepsilon \propto \frac{\alpha}{\theta_B}\sqrt{\frac{1}{N}} \qquad (1)$$

where $\theta_B$ is the Bragg angle with respect to which the strain is measured.

This analytical expression offers a way of interpolation to compare precision values for data acquired with different values of N and α. The constant of proportionality in Equation 1 depends on the signal profile, the detector geometry, and the choice of processing method. The first two factors remain mostly the same in all 4D-STEM datasets for strain mapping, namely circular diffraction disks and square-shaped pixels. The choice of processing method does affect the value of the numerical pre-factor, but variations in this factor between different methods is likely to have a far lesser influence on the precision of strain mapping compared to the dose or the aperture size used in the experiment. Hence, the use of different algorithms and their fine tuning can result in small improvements in precision, but for a large order of magnitude improvement, the most practical avenues are to use a higher dose or a more parallel beam. However, both these approaches for improving the precision are accompanied by a trade-off in spatial resolution, from the source-size contribution in the former case and the geometric (diffraction-limit) contribution in the latter case.

**Results**

We use a Si-SiGe multilayer sample oriented along the [110] zone axis to test the performance of the EWPC algorithm on 4D-STEM datasets taken on the EMPAD-G2 detector. We track the positions of the closest EWPC peaks in the $(1\bar{1}1)$ and $(\bar{1}11)$ directions and use their symmetric and anti-symmetric linear combinations as the basis vectors for the strain tensor in two dimensions. A virtual annular dark field (ADF) image of the multilayer sample is shown in Fig. 3(a), with the lower (higher) intensity regions corresponding to epitaxial Si (Si-Ge) layers. Strain maps calculated with the cepstral method are shown in Fig. 3(b) along with averaged line profile for the $\varepsilon_{zz}$ and $\varepsilon_{xx}$ components across the layers shown in Fig. 3(c). The datasets are acquired with a probe current of 1 nA and a dwell time of 100 us per diffraction pattern, allowing the full acquisition in less than 2 seconds. The high saturation current of the EMPAD-G2 detector (180 pA/pixel at 300 kV) and high brightness of the cold-field emission gun (C-FEG) used here are critical in enabling such fast acquisition. The $\varepsilon_{zz}$ map exhibits higher strain in the SiGe layer, consistent with its larger lattice spacing along the unconstrained growth direction. Naively, the $\varepsilon_{xx}$ map could be expected to appear featureless due to the in-plane epitaxial relationship. However, in TEM samples, elastic relaxation of thin sections can cause sample buckling that generates strain (Gibson & Treacy, 1984; Gibson et al., 1985). If tilts are large, this introduces an additional foreshortening artifact into the cepstral peak positions (Padgett et al., 2020). These effects are especially noticeable in the thinnest regions of the Si-SiGe multilayer, where the sample is bent and large mistilts lead to $\varepsilon_{xx}$ values that have a large deviation from the nominal zero, as shown in Fig. 3(c). Strain maps over the same region computed from datasets acquired with a higher dose are shown in Supplementary Fig. 1. While increased dose leads to a slight improvement in precision as evident from the smoother strain maps, the persistence of strain features in the $\varepsilon_{xx}$ map indicate that they stem from differences in dynamical diffraction between the two layers from the unavoidable thin-film relaxation coupled with a slight mistilt (Hÿtch & Minor, 2014). Although cross-correlation methods are not designed for small pixel-count detectors, we provide strain maps obtained using the cross-correlation method in py4DSTEM (Savitzky et al., 2021) in Supplementary Fig. 2 for comparison. While noisier as expected, the cross-correlation results also show the apparent strain features in $\varepsilon_{xx}$.

We also show a comparison of strain maps calculated using the EWPC method with EMPAD detector and that calculated using a commercial software and a 16M pixel CMOS camera in Supplementary Fig. 3. Systematic errors from thickness fringes and defects are more effectively suppressed in the EWPC strain map. The EMPAD's ability to record large currents at high speed and high DQE makes it possible to work with 2.5x lower dose and 5x faster acquisition compared to the low-saturation-current CMOS and MAPS detectors.

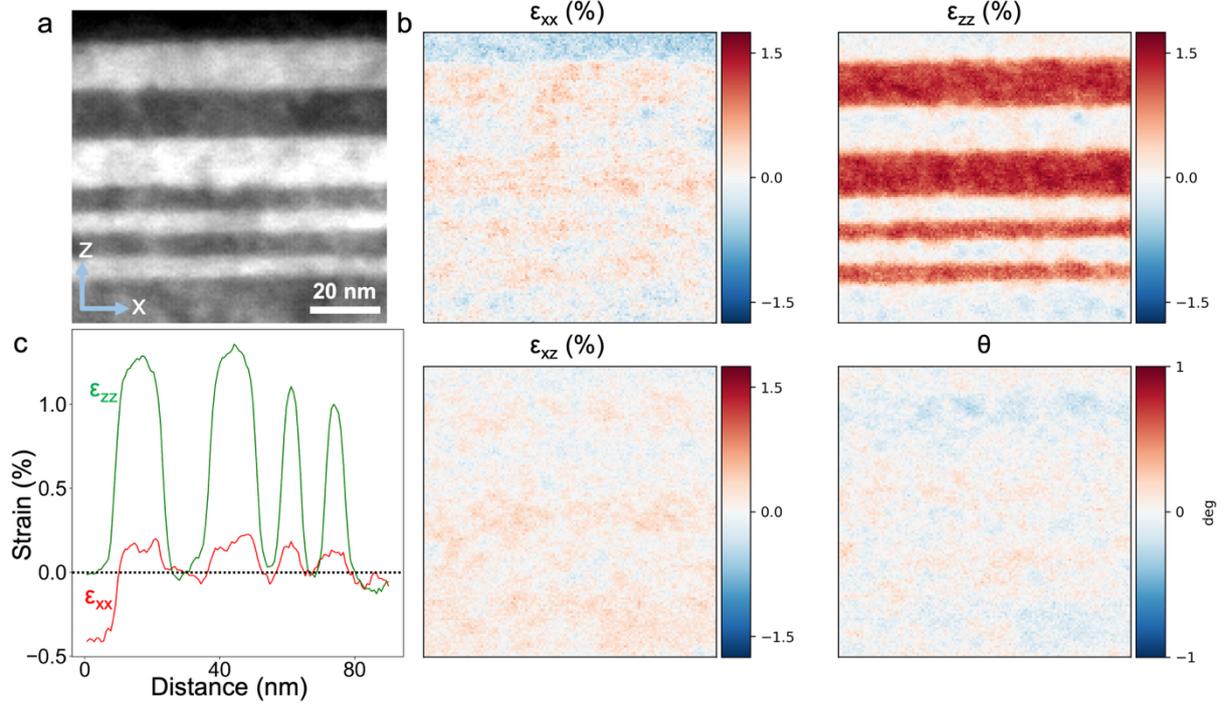

**Fig. 3. Cepstral strain mapping with 4D-STEM datasets acquired in under 2 seconds with the EMPAD-G2 detector.** (a) Virtual ADF image of the Si-SiGe heterostructure. (b) Strain maps calculated using the cepstral method. The $\varepsilon_{zz}$ map shows a higher strain in the SiGe layers as expected from the larger lattice spacing. The $\varepsilon_{xx}$ map would ideally be featureless as the layers are epitaxially connected but instead shows systematic errors across the Si-SiGe layers from dynamical diffraction and mistilt artifacts. Averaged line profile across the different layers for $\varepsilon_{xx}$ and $\varepsilon_{zz}$ map are also shown in (c).

In order to understand the influence of different experimental parameters on strain-mapping with EWPC, we performed systematic acquisition and analysis of 4D-STEM datasets with different values of probe semi-convergence angle, probe current and dwell time per pixel. The experiments here were done on a microscope with a lower brightness Schottky X-FEG (remotely accessed during the COVID lockdown) in comparison to the C-FEG source used for the experiments in Fig. 3. The results of this analysis are summarized in Fig. 4. The precision achievable for strain mapping is largely determined by the dose as shown in Fig. 4(a), where the precision is plotted as a function of dose (N) for different values of semi-convergence angle (α) on a log-log scale. The thickness of the sample region used to measure the strain mapping precision is estimated to be around 110 nm by comparing the PACBED pattern with a thickness series simulated with the Bloch wave method. The standard deviation in the measured strain over small regions (around a 10x10 pixel region) of the Si [110] sample is used as a measure of the local precision of our method. The mean of the precision of the diagonal components of the strain matrix is used as the data point with the standard deviation used as a rough estimate for the error bar. The precisions measured from same region of the sample shown in Supplementary Fig. 3 for both the EWPC/EMPAD and CMOS data are also marked (these two data points are at a different sample thickness to the colored data series). The dominant trend is a scaling of the precision with the Poisson noise as $\frac{1}{\sqrt{N}}$. The precision improves with increasing dose and follows the expected scaling at low doses (< 1pC). However, at higher doses, the precision plateaus out when systematic errors start to dominate over the Poisson noise. These systematic errors arise largely from the effects of the residual dynamical diffraction contrast that persist after the cepstral transform.

The data points for each convergence angle are fitted with $\sigma = \sqrt{\left(\frac{a}{\sqrt{N}}\right)^2 + b^2}$, where a and b are constants. The first term which represents the limit set by Poisson statistics, and the second term which represents other sources of noise are added in quadrature to obtain this functional form. Further, the values of the fit parameters a and b can be plotted as a function of the convergence angle to investigate the scaling of precision with the diffraction disk size and sources of systematic errors respectively. The details of this analysis shown in Supplementary Text 2 show that the precision has a linear dependence on α in the Poisson-noise limited regime as expected from Equation 1. However, increased precision from a smaller α comes with the cost of degradation of spatial resolution. This trade-off is shown in Fig. 4(b) where the precision is plotted as a function of the probe size on a log-log scale. Note that the probe size ($d_s$) values are not simply the diffraction-limited resolution for a given α, but also account for the effect of the finite source size. Given the additional coupling between source size and beam current, to achieve a 0.1% precision at 1 nm resolution, one would be better off using α = 2 mrad rather than 1.6 or 3 mrad.

Using the EWPC method, a measured precision of 0.09% and a resolution of 1 nm is achieved with a dose of 1 pC with α = 2 mrad, recorded with a dwell time of 10 ms. For a higher dose of 5 pC (500 pA, 10 ms dwell time), the precision attained is 0.06% for α = 1.6 mrad ($d_s$ = 2.5 nm) and 0.07% for α = 2 mrad ($d_s$ = 1.9 nm).

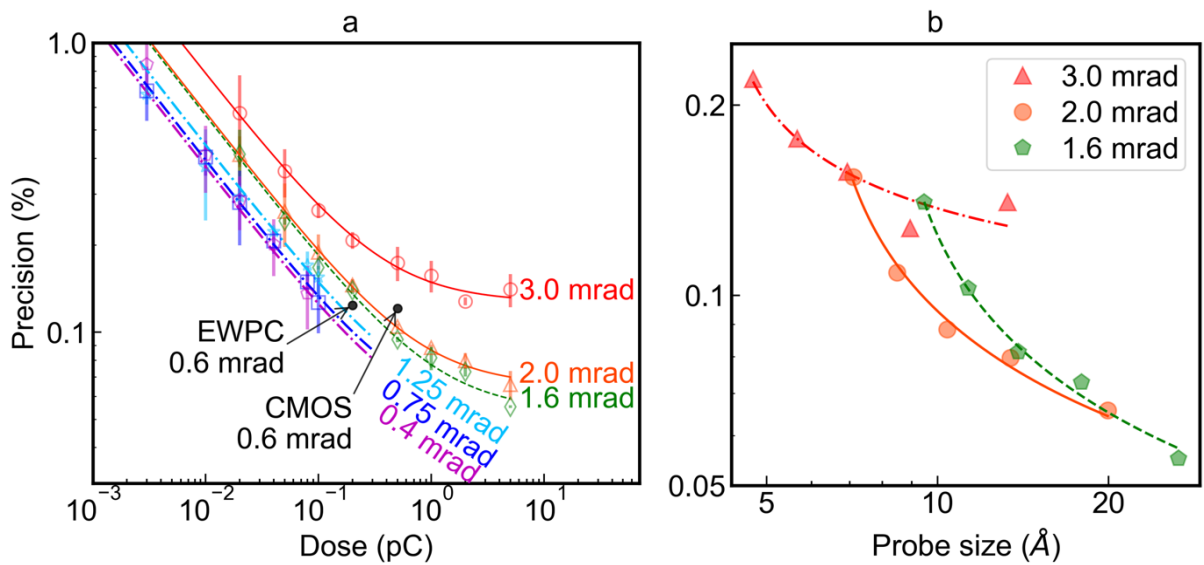

**Fig. 4. Measured local precision for strain mapping with different experimental conditions.** (a) Precision as a function of dose (N) on a log-log scale plotted for different values of probe semi-convergence angle (α). The data points are fit with a function of $\frac{1}{\sqrt{N}}$ from the Poisson noise added in quadrature with a constant from other noise sources. The precisions measured from same region of the data sets in Supplementary Fig. 3 for both the EWPC/EMPAD and CMOS data are also marked (different sample thickness to colored data series). (b) Precision as a function of the probe size on a log-log scale, showing the trade-off between higher precision and better resolution.

We also explored the effect of sample thickness on strain mapping by preparing a Si [110] sample with a gradual thickness variation as can be seen from the parallel thickness fringes in the virtual dark field (($1\bar{1}1$) reflection) image in Fig. 5 (a). The thickness varies gradually from about 80 nm at the top to about 170 nm at the bottom as determined by comparison with Bloch wave simulations shown in Supplementary Fig. 4. The dataset is taken with a 1 mrad semi-convergence angle and a dose of 1 pC. Since the sample is unstrained, the strain map should ideally be a flat, uniform distribution. However, the strain map calculated using the cepstral method in Fig. 5(b) shows alternating bands of higher (red) / lower (blue) strain varying at the same length scale as the thickness fringes in Fig. 5(a). As the sample should have nominally zero-strain, the mean of the measured strain can be used to represent the accuracy of the method, while the standard deviation describes the precision. These measurements can be carried out parallel to the thickness fringes to get the precision and accuracy as a function of thickness, with the results summarized in Figs. 5 (c) and (d) respectively. The variations in precision and accuracy as a function of thickness indicate that the performance of strain-mapping has a strong dependence on the sample region used for the measurement. This is to say that thickness and tilt variations between different samples still have larger variations in the reproducibility on the optimized measurements than the choice of processing method.

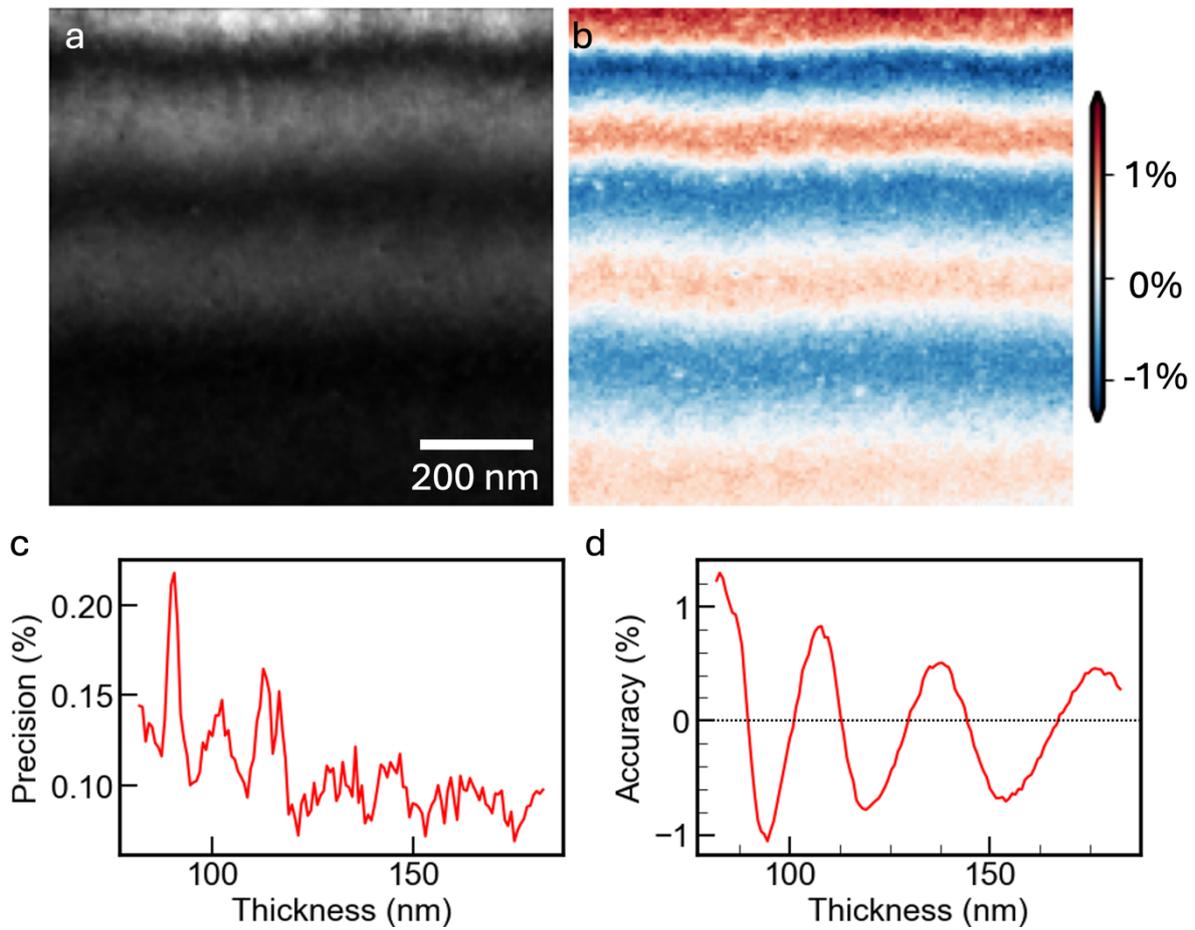

**Fig. 5. Strain mapping performance as a function of sample thickness on a Si [110] wedge.** The wedge gets gradually thicker towards the bottom of the figure. (a) Virtual dark field image of the wedge created with ($1\bar{1}1$) Bragg spot. (b) Strain map computed using the cepstral algorithm showing strain variations that mirror the thickness fringes observed in the dark field image. (c) Precision and (d) accuracy of the cepstral strain mapping plotted as a function of sample thickness measured from the strain map in (b).

*Impact of precession*

Precession electron diffraction (PED) is a technique that involves rocking the electron beam about the optic axis at an angle during acquisition, averaging over a range of incident angles. The resulting diffraction pattern is "pseudo-kinematical" with dynamical diffraction contrast largely suppressed. Naturally, this leads to less error in detecting disk shifts and hence improves the precision of strain mapping. As the beam is precessed, the Ewald sphere excites more reciprocal space points, resulting in additional higher-order diffraction spots appearing in the ZOLZ. With increasing number of diffraction spots to sample from, the cepstral method may be expected to become even more precise.

To study the impact of precession on strain mapping using EWPC, we acquired 4D-STEM datasets both i) with and ii) without precession (precession angle of ~1 degree) on a Si wedge sample oriented along the [110] zone axis. Figure 6(a) shows a direct comparison of the measured strain maps for a dose of 2 pC (200 pA beam current for a dwell time of 10 ms) and a semi-convergence angle of 2.5 mrad. A visual inspection of the strain maps clearly indicates that precession is highly effective in suppressing strain artifacts arising from thickness fringes and tilt variations (present at sample edge due to buckling). This improvement in precision is quantified in Figs. 6 (b, c) where the local and global precision respectively are plotted as a function of dose for two values of the semi-convergence angle (1.25 and 2.5 mrad). The datapoints are fit with the same functional form used in Fig. 4(a).

For low dose conditions at $\alpha$ =2.5 mrad, there is not much difference in the measured local precision with/without precession, likely because the Poisson noise is much larger than the systematic errors from dynamical diffraction contrast. However, for $\alpha$ =1.25 mrad and for higher dose conditions at $\alpha$ =2.5 mrad, there is a clear improvement in precision by a factor of 1.3-2x when precession is employed as can be seen in Fig. 6 (b). For a dose of 2 pC (dwell time of 10 ms/pixel), we obtain a precision of 0.03% (0.04%) at 2 nm spatial resolution, and 0.07% (0.1%) at 1 nm spatial resolution with (without) precession. The resolution reported above does not account for the slight degradation in spatial resolution during precession, caused by the coupling of aberrations and misalignments to the probe displacement.

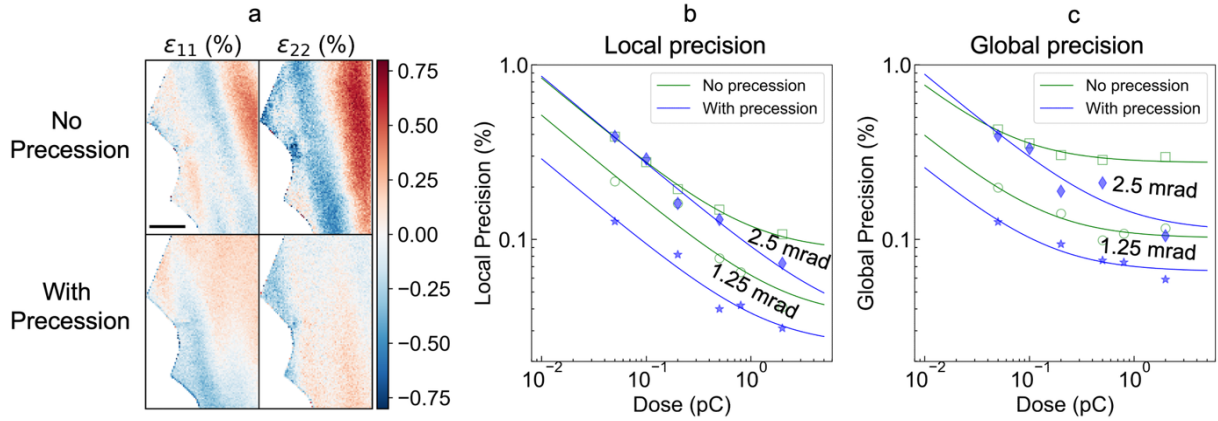

**Fig. 6. Influence of precession on strain mapping.** (a) Direct comparison of strain maps calculated from datasets taken with and without precession at the same experimental conditions (dose = 2 pC, α = 2.5 mrad). Plots comparing the change in (b) local precision and (c) global precision when precession is used. Solid blue markers and hollow green markers represent the datapoints with and without precession respectively. Scalebar is 200 nm.

The local precision metric cannot capture information about the variations in the calculated strain over regions with different diffraction conditions. The global precision metric which is the standard deviation in the measured strain over a large field of view is used to represent the effectiveness of the algorithm in handling such variations in the diffraction conditions. Since practical devices are expected to have a wide range of thickness variation as well as different material structures and interfaces across the device dimensions, the global precision might be a more apt metric for determining how well strain mapping methods might be expected to perform when applied to devices. In Fig. 6 (c), the global precision is plotted against the dose for the same datasets used to produce Fig. 6 (b). Overall, we observe a factor of 1.5-2.5x improvement in the global precision when precession is used, with the greatest impact for the 2.5 mrad semi-convergence angle at high doses. This is expected since the dynamical diffraction contrast would be more pronounced in larger diffraction disks, and precession is able to suppress artifacts arising from this contrast.

### *Impact of energy filtering*

Zero-loss energy filtering is another technique often used to improve the precision of strain measurement in relatively thick samples (where the inelastic contribution accounts for a considerable portion of the total scattered intensity). Removing the inelastic background in the diffraction pattern improves the sharpness of the diffraction spots. Hähnel et al., 2012 showed that energy filtering improved the strain mapping precision in thick Si samples (~600 nm) by more than 50%. However, for a 200 nm thick Si sample, Favia et al., 2011 reported no improvement in the precision with energy filtering, despite an improved contrast in the recorded diffraction patterns. We studied whether energy filtering of the diffraction pattern has an influence on strain mapping with EWPC by using an energy window of width 10 eV centered at the zero-loss peak to filter out the inelastically scattered electrons. Crudely, the inelastically scattered electrons have the effect of an isotropic blur of the diffraction pattern features. Hence, in cepstral space, inelastic scattering would have the effect of multiplication with an envelope. Since the blur in diffraction space has a small width, the envelope in the cepstral space decays very slowly away from the center. Hence, the cepstral space peaks near the center that are typically used for strain mapping are not strongly affected by the inelastic background in the diffraction pattern. Figure 7 compares strain maps obtained with and without energy filtering for the same sample region as in Fig. 6, showing that energy filtering does not have any considerable impact on strain mapping with EWPC as long as the samples are not too thick (roughly less than one inelastic mean free path).

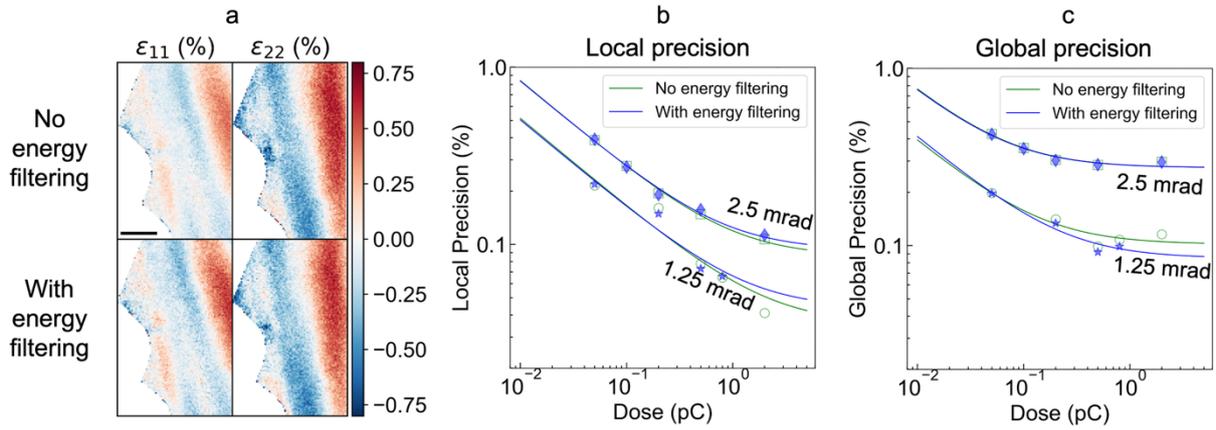

**Fig. 7. Influence of zero-loss energy filtering on strain mapping.** (a) Direct comparison of strain maps calculated from data taken with and without energy filtering at the same experimental conditions (dose = 2 pC, α = 2.5 mrad). Plots comparing the change in (b) local precision and (c) global precision as a function of dose for different datasets taken with and without energy filtering. Solid blue markers and hollow green markers represent the datapoints with and without energy filtering respectively. Very little difference is observed between the filtered and unfiltered results. Scalebar is 200 nm.

However, in instruments where the pixelated detector lies at the end of the spectrometer, energy-loss dependent aberrations induced by the spectrometer can distort the diffraction pattern. Therefore, zero-loss energy filtering is recommended when the spectrometer is used, whereas it is not necessary otherwise.

# Discussion

The cepstral algorithm is designed to maximally utilize the dose used in the experiment and work effectively on data collected on small pixel count detectors, where each diffracted disk is localized over only a few pixels. This is achieved by taking the Fourier transform of the diffraction pattern, which enfolds the information from all diffracted disks and makes use of all the scattered electron counts within the Zeroth Order Laue Zone. In contrast, typical strain mapping algorithms which use edge detection or cross-correlation methods focus on only a few diffracted spots, hence not fully making use of the scattered electron counts. Taking a logarithm of the diffraction pattern before the Fourier transform helps in reducing the impact of the relative intensity difference between the different scattered disks and gives more importance to the relative positions of the disks or the periodicity in the pattern, giving increased robustness against artifacts from dynamical diffraction contrast arising due to thickness and tilt variations. Since each cepstral spot is derived from the periodicity of several diffraction spots, one can expect the effects from the dynamical diffraction contrast to at least partially average out. For example, in the case of thickness fringes, the intensities of different diffracted beams have different periods of oscillation (depending on the extinction length and the excitation error), and hence the different phases of the diffracted beams cancel out at least partially on average, reducing the influence of thickness variations on the peak positions of the cepstral spots. This can be compared to how a dark field image from a particular Bragg reflection has strong thickness fringes whereas an annular dark field image which averages over a large number of diffracted beams has minimal diffraction contrast. However, the logarithm also reduces the intensity difference between diffraction peaks and the background, so can overweight the influence of noise fluctuations from the diffuse scattering in low-dose diffraction patterns. A pre-filtering step to suppress such noise contributions is recommended if EWPC is to be used for low-dose analysis.

Moreover, the EWPC transform is particularly suited for strain mapping applications where a high throughput is desired, because of multiple reasons - i) The method makes use of all the diffracted beams within the Zeroth Order Laue Zone (ZOLZ) and is highly dose efficient. Because of the high dose-efficiency, a relatively short dwell time of 0.1-10 ms/pixel can be used. ii) The periodicity of the different diffraction spots in a direction is now encoded in a single EWPC peak. Instead of locating the precise positions of all the diffracted disks along a particular direction, it suffices to calculate the precise position of just a single EWPC spot, saving computation time. iii) Since the EWPC method works well even with a few number of detector pixels, this enables working with compact data sets that can be easily handled. Since the computation time scales up with the number of pixels, this also considerably speeds up the processing time. iv) The EWPC algorithm can be completely automated and does not require tuning of any parameters depending on the dataset, other than deciding which EWPC peaks to track. This absence of a need to optimize fitting parameters is often not the case for other strain mapping algorithms, particularly the edge detection algorithms. However, some care is needed in choosing which EWPC peaks to track – peaks that arise from forbidden reflections like the double-scattering {002} peaks in silicon should be avoided as they have a very strong thickness dependence.

## Summary

The key ingredient for high-throughput, high-precision strain mapping is collecting maximum dose on the sample in the shortest time, making large saturation current pixel array detectors well suited for this purpose. However, due to their small pixel counts, traditional strain mapping algorithms that rely on large pixel counts perform poorly on data acquired on pixel array detectors. In this study, we have demonstrated the utility of the cepstral algorithm and the EMPAD detectors in achieving high precision strain maps even with only a few pixels per diffraction disk. This combination offers fast acquisition and processing times, making the technique potentially well suited for high-throughput applications. We discussed the optimal experimental conditions for the EWPC method and demonstrated the robustness of the technique to diffraction artifacts from tilt and thickness in comparison to industry grade software. We obtain a precision of 0.09% and a resolution of 1 nm with a dose of 1 pC with α = 2 mrad, recorded with a dwell time of 10 ms.

Comparing the datasets obtained at different experimental conditions, we showed how the precision scales as a function of dose and the semi-convergence angle, and the trade-off between higher precision and better resolution. As expected, we found that the use of precession results in an increased robustness to diffraction artifacts, and an improved precision reaching 0.03% at 2 nm resolution with a 2 pC dose. Our results also show that energy filtering does not have a considerable impact on the precision of the strain mapping using EWPC for thin samples. A direct comparison of different strain mapping algorithms (MacLaren et al., 2021) is often quite difficult due to incomplete reports of the sample and experiment details in literature. Variations in the sample thickness and tilt often have a much larger influence on the accuracy and precision of strain mapping than the differences between different algorithms, and hence strain mapping on a standard sample with a full report of the probe current, dwell time, resolution and sample thickness might be a good starting point if different algorithms need to be compared quantitatively. For this study, we used a Si-SiGe multilayer sample and Si wedges imaged along the [110] zone axis, which could be good standard samples for such tests due to its universal use in semiconductor devices, also making it directly relevant to industry applications.

**Code Availability**

Both the MATLAB and Python versions of the code for cepstral analysis are made freely available and can be accessed at the following links:

MATLAB - https://github.com/muller-group-cornell/PC-STEM

Python - https://github.com/muller-group-cornell/Cepstral_analysis_Python

**Acknowledgements**

HKP was supported by the Platform for the Accelerated Realization, Analysis, and Discovery of Interface Materials (PARADIM) Innovation Platform by the National Science Foundation (NSF) grant DMR-2039380. DY was supported by the Center for Alkaline-based Energy Solutions, an Energy Frontier Research Center funded by the Department of Energy, Office of Science, BES under Award # DE-SC0019445. ZB was supported by the NSF under Grant # PHY-1549132, the Center for Bright Beams. YTS was supported by the Department of Defense, Air Force Office of Scientific Research under award FA9550-18-1-0480. This work made use of the Cornell Center for Materials Research (CCMR) facilities supported by NSF (DMR-2039380), NSF-MRI-1429155 and NSF (DMR-1539918). The authors thank John Grazul, Phil Carubia, Steven Zeltmann, Mariena Silvestry Ramos and Malcolm Thomas for technical support and maintenance of the electron microscopy facilities. The authors thank Peter Tiemeijer (Thermo Fisher Scientific) for his support in setting up the energy filter for the experiments reported in this paper. We acknowledge Jiangtao Zhu at Eurofins Nanolab Technologies for providing the Si-SiGe sample and Malcolm Thomas and Shake Karapetyan for help with sample preparation. We also thank Steven Zeltmann and Prof. Andrew Minor for help with our questions regarding py4DSTEM and previously published work.

# References

ALLEN, L. J., D'ALFONSO, A. J. & FINDLAY, S. D. (2015). Modelling the inelastic scattering of fast electrons. *Ultramicroscopy* **151**, 11–22.

BOGERT B. P., HEALY M. J. R., & TUKEY JOHN W. (1963). The Quefrency Alanysis of Time Series for Echoes: Cepstrum, Pseudo-Autocovariance, Cross-Cepstrum and Saphe Cracking. *Proceedings of the Symposium on Time Series Analysis* 209–243.

CHAPMAN, J. N., BATSON, P. E., WADDELL, E. M. & FERRIER, R. P. (1978). The direct determination of magnetic domain wall profiles by differential phase contrast electron microscopy. *Ultramicroscopy* **3**, 203–214.

CHEN, Z., JIANG, Y., SHAO, Y. T., HOLTZ, M. E., ODSTRČIL, M., GUIZAR-SICAIROS, M., HANKE, I., GANSCHOW, S., SCHLOM, D. G. & MULLER, D. A. (2021). Electron ptychography achieves atomic-resolution limits set by lattice vibrations. *Science* **372**, 826–831.

COOPER, D., BERNIER, N. & ROUVIÈRE, J. L. (2015). Combining 2 nm Spatial Resolution and 0.02% Precision for Deformation Mapping of Semiconductor Specimens in a Transmission Electron Microscope by Precession Electron Diffraction. *Nano Letters* **15**, 5289–5294.

CUEVA, P., PADGET, E. & MULLER, D. A. (2018). A Natural Basis for Unsupervised Machine Learning on Scanning Diffraction Data. *Microscopy and Microanalysis* **24**, 490–491.

FAVIA, P., BARGALLO GONZALES, M., SIMOEN, E., VERHEYEN, P., KLENOV, D. & BENDER, H. (2011). Nanobeam Diffraction: Technique Evaluation and Strain Measurement on Complementary Metal Oxide Semiconductor Devices. *Journal of The Electrochemical Society* **158**, H438.

FORBES, B. D., MARTIN, A. V., FINDLAY, S. D., D'ALFONSO, A. J. & ALLEN, L. J. (2010). Quantum mechanical model for phonon excitation in electron diffraction and imaging using a Born-Oppenheimer approximation. *Physical Review B - Condensed Matter and Materials Physics* **82**, 104103–104103.

GHANBARI, R., KP, H., PATEL, K., ZHOU, H., ZHOU, T., LIU, R., WU, L., KHANDELWAL, A., CRUST, K. J., HAZRA, S., CARROLL, J., MEYERS, C. J. G., WANG, J., PROSANDEEV, S., QIAO, H., KIM, Y.-H., NABEI, Y., CHI, M., SUN, D., BALKE, N., HOLT, M., GOPALAN, V., SPANIER, J. E., MULLER, D. A., BELLAICHE, L., HWANG, H. Y. & XU, R. (2025). Strain-induced lead-free morphotropic phase boundary. *Nature Communications* **16**, 7766.

GIBSON, J. M., HULL, R., BEAN, J. C. & TREACY, M. M. J. (1985). Elastic relaxation in transmission electron microscopy of strained-layer superlattices. *Applied Physics Letters* **46**, 649–651.

GIBSON, J. M. & TREACY, M. M. J. (1984). The effect of elastic relaxation on the local structure of lattice-modulated thin films. *Ultramicroscopy* **14**, 345–349.

HÄHNEL, A., REICHE, M., MOUTANABBIR, O., BLUMTRITT, H., GEISLER, H., HÖNTSCHEL, J. & ENGELMANN, H. J. (2012). Improving Accuracy and Precision of Strain Analysis by Energy-Filtered Nanobeam Electron Diffraction. *Microscopy and Microanalysis* **18**, 229–240.

HAN, Y., NGUYEN, K. X., CAO, M., CUEVA, P. D., XIE, S., TATE, M. W., PUROHIT, P., GRUNER, S. M., PARK, J. & MULLER, D. A. (2018). Strain Mapping of Two-Dimensional Heterostructures with Sub-Picometer Precision. *Nano Letters* **18**, 3746–3751.

HARIKRISHNAN, K. P., YOON, D., SHAO, Y.-T., MELE, L., MITTERBAUER, C. & MULLER, D. (2021). Dose-efficient strain mapping with high precision and throughput using cepstral transforms on 4D-STEM data. *Microscopy and Microanalysis* **27**, 1994–1996.

HSIAO, H. W., FENG, R., NI, H., AN, K., POPLAWSKY, J. D., LIAW, P. K. & ZUO, J. M. (2022). Data-driven electron-diffraction approach reveals local short-range ordering in CrCoNi with ordering effects. *Nature Communications 2022 13:1* **13**, 1–9.

HŸTCH, M. J. & MINOR, A. M. (2014). Observing and measuring strain in nanostructures and devices with transmission electron microscopy. *MRS Bulletin* **39**, 138–146.

JIANG, Y., CHEN, Z., HAN, Y., DEB, P., GAO, H., XIE, S., PUROHIT, P., TATE, M. W., PARK, J., GRUNER, S. M., ELSER, V. & MULLER, D. A. (2018). Electron ptychography of 2D materials to deep sub-ångström resolution. *Nature* **559**, 343–349.

KAVLE, P., ROSS, A. M., KP, H., MEISENHEIMER, P., DASGUPTA, A., YANG, J., LIN, C.-C., PAN, H., BEHERA, P., PARSONNET, E., HUANG, X., ZORN, J. A., SHAO, Y.-T., DAS, S., LIU, S., MULLER, D. A., RAMESH, R., CHEN, L.-Q. & MARTIN, L. W. (2024). Highly Responsive Polar Vortices in All-Ferroelectric Heterostructures. *Advanced Materials* **36**, 2410146.

KP, H., XU, R., PATEL, K., CRUST, K. J., KHANDELWAL, A., ZHANG, C., PROSANDEEV, S., ZHOU, H., SHAO, Y.-T., BELLAICHE, L., HWANG, H. Y. & MULLER, D. A. (2025). Electron ptychography reveals a ferroelectricity dominated by anion displacements. *Nature Materials* 1–8.

LI, X., ZHENG, S. Q., EGAMI, K., AGARD, D. A. & CHENG, Y. (2013). Influence of electron dose rate on electron counting images recorded with the K2 camera. *Journal of Structural Biology* **184**, 251–260.

MACLAREN, I., DEVINE, E., GERGOV, H., PATERSON, G., HARIKRISHNAN, K. P., SAVITZKY, B., OPHUS, C., YUAN, R., ZUO, J.-M., FORSTER, K., KOBE, G., KOPPANY, E., MCCLYMONT, K., NARENDRAN, A. & RILEY, D. (2021). Comparing different software packages for the mapping of strain from scanning precession diffraction data. *Microscopy and Microanalysis* **27**, 2–5.

MACLAREN, I., FRUTOS-MYRO, E., MCGROUTHER, D., MCFADZEAN, S., WEISS, J. K., COSART, D., PORTILLO, J., ROBINS, A., NICOLOPOULOS, S., NEBOT DEL BUSTO, E. & SKOGEBY, R. (2020). A comparison of a direct electron detector and a high-speed video camera for a scanning precession electron diffraction phase and orientation mapping. *Microscopy and Microanalysis* **3**, 1080–1080.

MCMULLAN, G., FARUQI, A. R., CLARE, D. & HENDERSON, R. (2014). Comparison of optimal performance at 300 keV of three direct electron detectors for use in low dose electron microscopy. *Ultramicroscopy* **147**, 156–163.

MIDGLEY, P. A. & EGGEMAN, A. S. (2015). Precession electron diffraction - A topical review. *IUCrJ* **2**, 126–136.

MÜLLER, K., ROSENAUER, A., SCHOWALTER, M., ZWECK, J., FRITZ, R. & VOLZ, K. (2012). Strain measurement in semiconductor heterostructures by scanning transmission electron microscopy. *Microscopy and Microanalysis* **18**, 995–1009.

NELLIST, P. D., MCCALLUM, B. C. & RODENBURG, J. M. (1995). Resolution beyond the 'information limit' in transmission electron microscopy. *Nature* **374**, 630–632.

NGUYEN, K. X., ZHANG, X. S., TURGUT, E., CAO, M. C., GLASER, J., CHEN, Z., STOLT, M. J., CHANG, C. S., SHAO, Y. T., JIN, S., FUCHS, G. D. & MULLER, D. A. (2022). Disentangling Magnetic and Grain Contrast in Polycrystalline FeGe Thin Films Using Four-Dimensional Lorentz Scanning Transmission Electron Microscopy. *Physical Review Applied* **17**, 1.

OPHUS, C. (2019). Four-Dimensional Scanning Transmission Electron Microscopy (4D-STEM): From Scanning Nanodiffraction to Ptychography and Beyond. *Microscopy and Microanalysis* 563–582.

PADGETT, E., HOLTZ, M. E., CUEVA, P., SHAO, Y. T., LANGENBERG, E., SCHLOM, D. G. & MULLER, D. A. (2020). The exit-wave power-cepstrum transform for scanning nanobeam electron diffraction: robust strain mapping at subnanometer resolution and subpicometer precision. *Ultramicroscopy* **214**, 112994.

PADGETT, E., HOLTZ, M. E., KONGKANAND, A. & MULLER, D. A. (2023). Strain Relaxation in Core-Shell Pt-Co Catalyst Nanoparticles. http://arxiv.org/abs/2305.18686 (Accessed April 24, 2025).

PATTERSON, A. L. (1934). A Fourier Series Method for the Determination of the Components of Interatomic Distances in Crystals. *Physical Review* **46**, 372–372.

PEKIN, T. C., GAMMER, C., CISTON, J., MINOR, A. M. & OPHUS, C. (2017). Optimizing disk registration algorithms for nanobeam electron diffraction strain mapping. *Ultramicroscopy* **176**, 170–176.

PHILIPP, H. T., TATE, M. W., SHANKS, K. S., MELE, L., PEEMEN, M., DONA, P., HARTONG, R., VAN VEEN, G., SHAO, Y. T., CHEN, Z., THOM-LEVY, J., MULLER, D. A. & GRUNER, S. M. (2022). Very-High Dynamic Range, 10,000 Frames/Second Pixel Array Detector for Electron Microscopy. *Microscopy and Microanalysis* **28**, 425–440.

RAUCH, E. F., PORTILLO, J., NICOLOPOULOS, S., BULTREYS, D., ROUVIMOV, S. & MOECK, P. (2010). Automated nanocrystal orientation and phase mapping in the transmission electron microscope on the basis of precession electron diffraction. *Zeitschrift fur Kristallographie* **225**, 103–109.

SAVITZKY, B. H., ZELTMANN, S. E., HUGHES, L. A., BROWN, H. G., ZHAO, S., PELZ, P. M., PEKIN, T. C., BARNARD, E. S., DONOHUE, J., RANGEL DACOSTA, L., KENNEDY, E., XIE, Y., JANISH, M. T., SCHNEIDER, M. M., HERRING, P., GOPAL, C., ANAPOLSKY, A., DHALL, R., BUSTILLO, K. C., ERCIUS, P., SCOTT, M. C., CISTON, J., MINOR, A. M. & OPHUS, C. (2021). py4DSTEM: A Software Package for Four-Dimensional Scanning Transmission Electron Microscopy Data Analysis. *Microscopy and Microanalysis* **27**, 712–743.

SHAO, Y. T., YUAN, R., HSIAO, H. W., YANG, Q., HU, Y. & ZUO, J. M. (2021). Cepstral scanning transmission electron microscopy imaging of severe lattice distortions. *Ultramicroscopy* **231**.

STROPPA, D. G., MEFFERT, M., HOERMANN, C., ZAMBON, P., BACHEVSKAYA, D., REMIGY, H., SCHULZE-BRIESE, C. & PIAZZA, L. (2023). From STEM to 4D STEM: Ultrafast Diffraction Mapping with a Hybrid-Pixel Detector. *Microscopy Today* **31**, 10–14.

SUN, Z., BARAISSOV, Z., PORTER, R. D., SHPANI, L., SHAO, Y.-T., OSEROFF, T., THOMPSON, M. O., MULLER, D. A. & LIEPE, M. U. (2023). Smooth, homogeneous, high-purity Nb3Sn superconducting RF resonant cavity by seed-free electrochemical synthesis. *Superconductor Science and Technology* **36**, 115003.

TATE, M. W., PUROHIT, P., CHAMBERLAIN, D., NGUYEN, K. X., HOVDEN, R., CHANG, C. S., DEB, P., TURGUT, E., HERON, J. T., SCHLOM, D. G., RALPH, D. C., FUCHS, G. D., SHANKS, K. S., PHILIPP, H. T., MULLER, D. A. & GRUNER, S. M. (2016). High Dynamic Range Pixel Array Detector for Scanning Transmission Electron Microscopy. *Microscopy and Microanalysis* **22**, 237–249.

VINCENT, R. & MIDGLEY, P. A. (1994). Double conical beam-rocking system for measurement of integrated electron diffraction intensities. *Ultramicroscopy* **53**, 271–282.

YOO, T., HERSHKOVITZ, E., YANG, Y., DA CRUZ GALLO, F., MANUEL, M. V. & KIM, H. (2024). Unsupervised machine learning and cepstral analysis with 4D-STEM for characterizing complex microstructures of metallic alloys. *npj Computational Materials* **10**, 1–10.

YOON, D., K.P, H., SHAO, Y.-T. & MULLER, D. A. (2022). High-Speed, High-Precision, and High-Throughput Strain Mapping with Cepstral Transformed 4D-STEM Data. *Microscopy and Microanalysis* **28**, 796–798.

YUAN, R., ZHANG, J. & ZUO, J. M. (2019). Lattice strain mapping using circular Hough transform for electron diffraction disk detection. *Ultramicroscopy* **207**, 112837.

ZAMBON, P., BOTTINELLI, S., SCHNYDER, R., MUSARRA, D., BOYE, D., DUDINA, A., LEHMANN, N., DE CARLO, S., RISSI, M., SCHULZE-BRIESE, C., MEFFERT, M., CAMPANINI, M., ERNI, R. & PIAZZA, L. (2023). KITE: High frame rate, high count rate pixelated electron counting ASIC for 4D STEM applications featuring high-Z sensor. *Nuclear Instruments and Methods in Physics Research Section A: Accelerators, Spectrometers, Detectors and Associated Equipment* **1048**, 167888.

ZELTMANN, S. E., MÜLLER, A., BUSTILLO, K. C., SAVITZKY, B., HUGHES, L., MINOR, A. M. & OPHUS, C. (2020). Patterned probes for high precision 4D-STEM bragg measurements. *Ultramicroscopy* **209**. http://arxiv.org/abs/1907.05504.

ZHANG, X., PADGETT, E., ZHU, L., BUHRMAN, R. & MULLER, D. (2020). A Robust Basis for Grain Identification in Polycrystalline Thin Film Devices Using Cepstrum Transforms of 4D-STEM Diffraction Pattern. *Microscopy and Microanalysis* **26**, 1620–1622.

ZUO, J. M. & MABON, J. C. (2004). Web-Based Electron Microscopy Application Software: Web-EMAPS. *Microscopy and Microanalysis* **10**, 1000–1001.

ZUO, J. M. & SPENCE, J. C. H. (2016). Advanced transmission electron microscopy: Imaging and diffraction in nanoscience. *Advanced Transmission Electron Microscopy: Imaging and Diffraction in Nanoscience* 1–729.

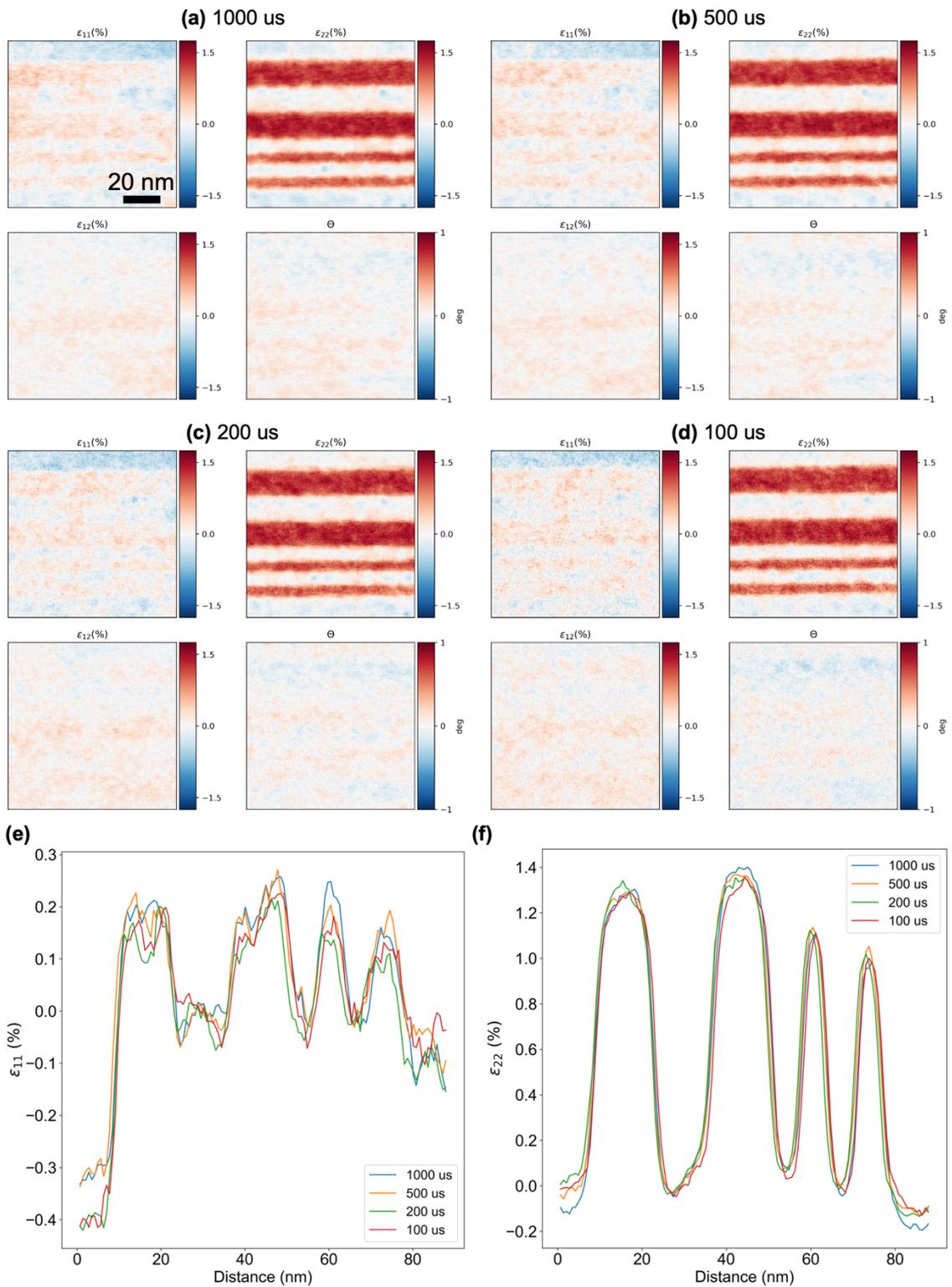

**Supplementary Fig. 1. Strain mapping of the Si-SiGe sample from datasets acquired at different doses by changing the dwell time per pixel -** (a) 1000 us, (b) 500 us, (c) 200 us, (d) 100 us. (e, f) Profiles of the $\varepsilon_{11}$ and $\varepsilon_{22}$ strain maps respectively measured across the multilayer sample.

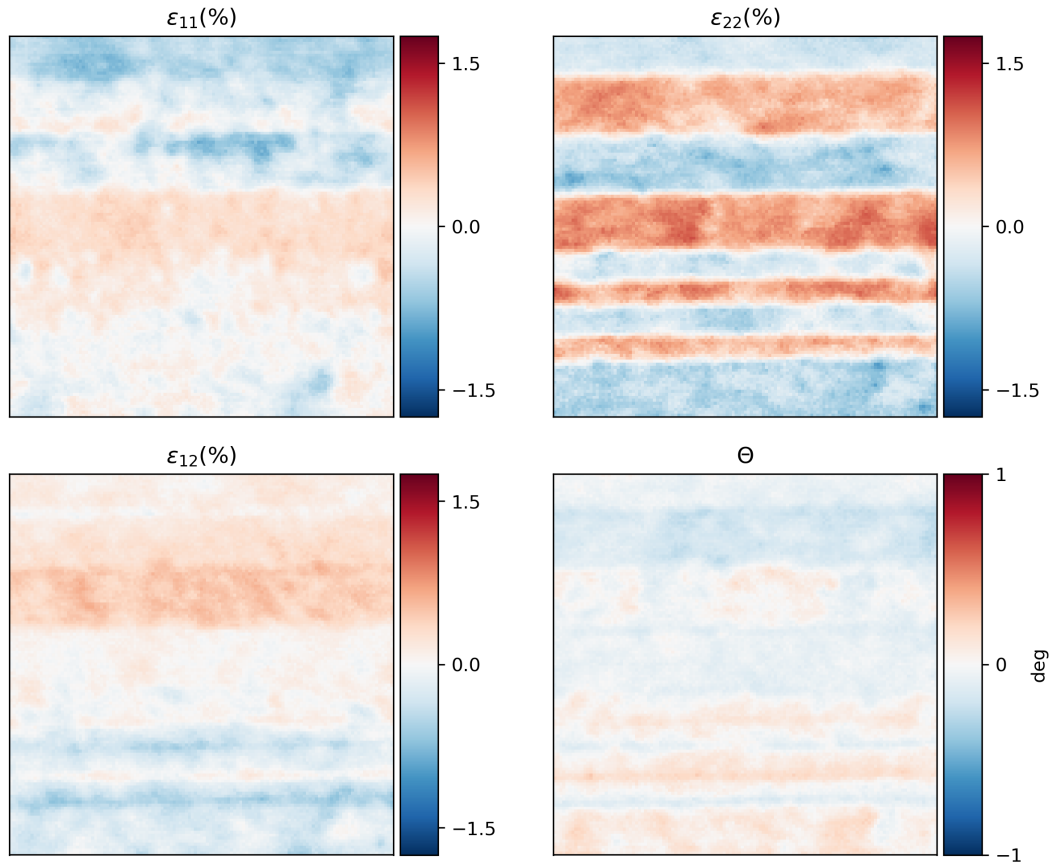

**Supplementary Fig. 2. Strain maps obtained on the Si-SiGe sample with the cross-correlation method in py4DSTEM**. The strain maps are calculated from the same dataset used to generate the cepstral strain map shown in Fig. 3

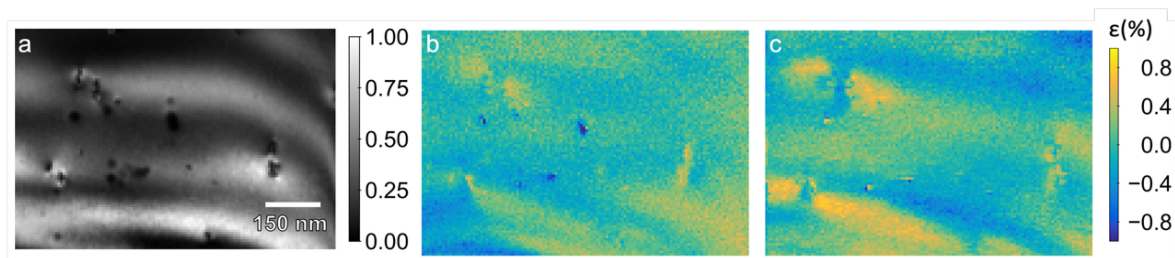

**Supplementary Fig. 3. Robustness of EWPC strain mapping to thickness artefacts in comparison to industry grade software**. (a) Virtual dark field image from the $(1\bar{1}1)$ reflection of a silicon sample oriented along the [110] zone axis. Thickness fringes and defects over the field of view can be clearly seen. Strain maps over the same region obtained using the (b) EWPC method (dose=0.2 pC, 10 ms/pixel) with data collected on the 128 x 128 pixel EMPAD detector and (c) Commercial Strain Analysis software with data collected on a 16M pixel CMOS camera (dose=0.5 pC, 50 ms/pixel). A semi-convergence angle of 0.6 mrad is used in both cases. Radiation damage marks from previous x-ray experiments were used as markers to align the fields of view for the two different experiments.

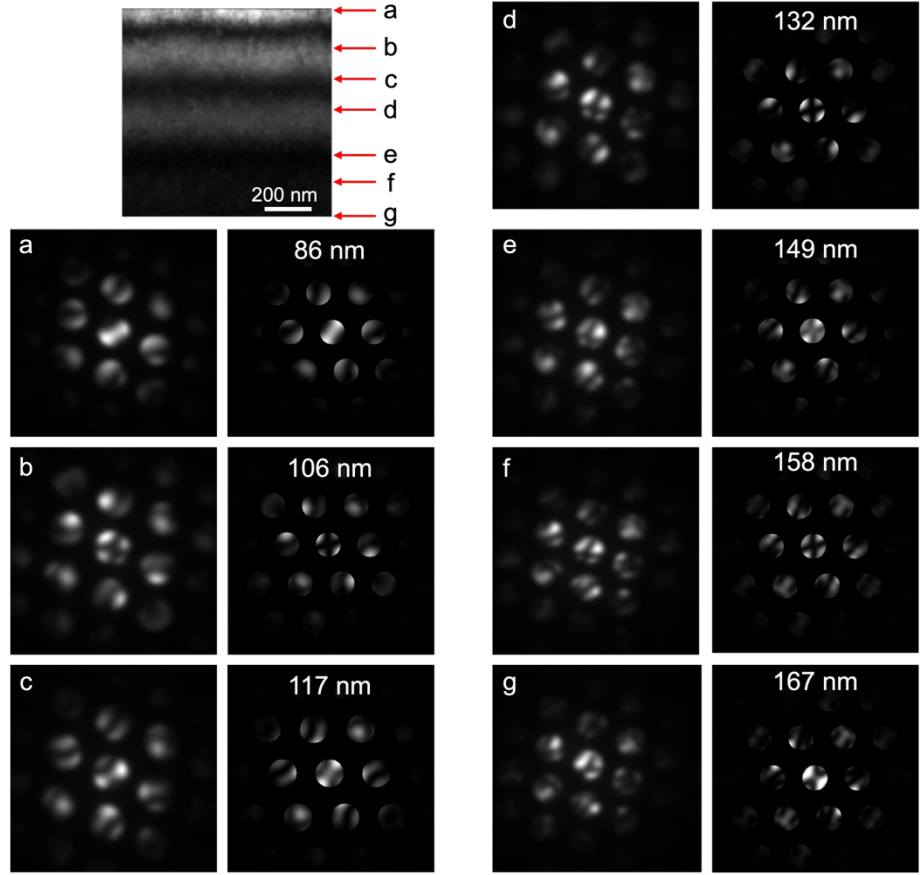

**Supplementary Fig. 4. Determination of sample thickness by comparison with Bloch wave simulations.** The thickness of different regions in the wedge were determined by comparing the nanobeam diffraction patterns with a thickness series of Bloch wave simulations. A linear fit was then used to determine the thickness range used in Fig. 5.

**<u>Supplementary Text 1 – Peak Finding Methods</u>**

We used the Nelder-Mead simplex algorithm on the Fourier transform function defined to have a discrete domain in reciprocal space and a continuous range in cepstral space to find peak positions in the EWPC pattern with sub-pixel precision. This is mathematically equivalent to using the Whittaker-Shannon interpolation (Equation S1) on the discrete EWPC pattern calculated using a discrete Fourier Transform. In a practical sense, the only differences would arise from the windowing function used in the Fast Fourier Transform and the tolerance values passed to the simplex algorithm. As expected, both these methods take roughly the same amount of time (around 36 ms for a single peak fit on a standard PC in MATLAB). For speedup in the peak fit, we explored the Lanczos interpolation method (Equation S2a). The Lanczos kernel (Equation S2b) is very similar to the sinc function kernel used in the Whittaker-Shannon interpolation, but is windowed with another sinc function which has its first zero at the filter size parameter r and is set to zero outside r.

For a discrete function f(i,j) with i,j $\in \mathbb{Z}$ in 2 dimensions defined over a m × n pixel region:

$$\text{Whittaker Shannon interpolation: } f(x, y) = \sum_{i=1}^{m} \sum_{j=1}^{n} f(i,j) \text{sinc}(x - i) \text{sinc}(y - j) \quad (S1)$$

$$\text{Lanczos interpolation: } f(x,y) = \sum_{i=|x|-r+1}^{|x|+r} \sum_{j=|y|-r+1}^{|y|+r} f(i,j) L(x - i) L(y - j) \quad (S2a)$$

$$\text{where the Lanczos kernel } L(x) = \begin{cases} \text{sinc}(x) \text{sinc}\left(\frac{x}{r}\right) & \text{if } -r<x<r, \\ 0 & \text{otherwise.} \end{cases} \quad (S2b)$$

Using a filter parameter value r = 10 gives results reasonably close to the value from the Whittaker-Shannon interpolation and allows a speedup by a factor of more than 2, with a single peak fit taking about 15 ms in MATLAB. These times can be reduced using compiled code, which would be important for live processing.

**Supplementary Text 2 - Dependence of fit parameters on convergence angle**

In Supplementary Fig. 5, we plot how the parameters *a* and *b* used in the fit in Fig. 4 (a) vary as a function of the convergence angle. The functional form of the fit $\delta\varepsilon = \sqrt{\left(\frac{a}{\sqrt{N}}\right)^2 + b^2}$ can be compared to the empirical expression $\delta\varepsilon \propto \frac{\alpha}{\theta_B} \sqrt{\frac{1}{N}}$ (Equation 1).

The empirical model predicts a linear dependence of the precision on the semi-convergence angle, which is roughly what we observe from the plot of *a* vs α in Supplementary Fig. 5. The parameter *b* is expected to tell us about the systematic errors in the experiment. The value of *b* increases slowly as α is increased from 0.4 mrad to 1.6 mrad, and then rises rapidly as it is further increased. This indicates that the parameter *b* might be representing the errors picked up from the dynamical diffraction contrast within the disks.

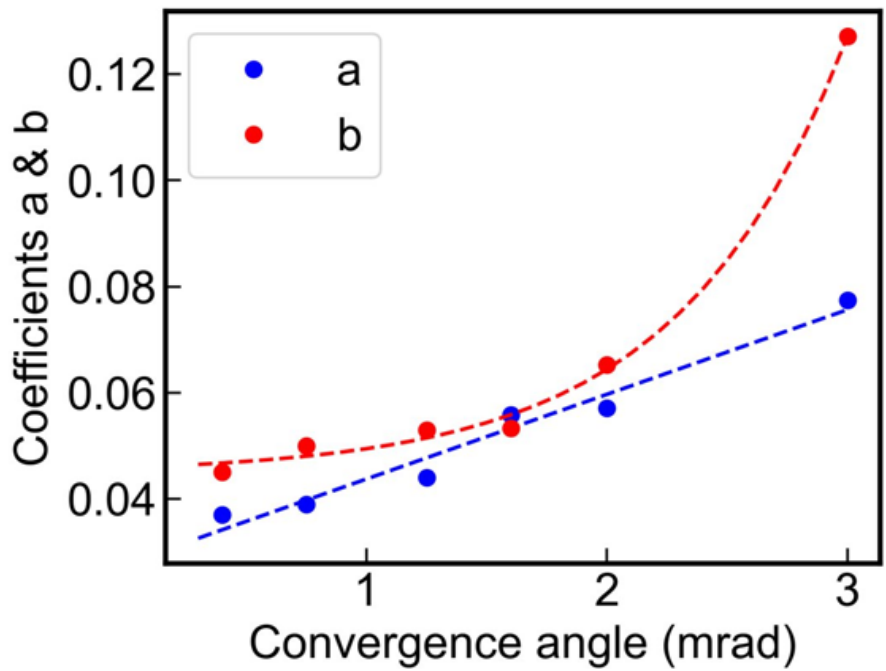

**Supplementary Fig. 5**. Plot of the parameters a and b used for the curve of the precision, $\sigma = \sqrt{\left(\frac{a}{\sqrt{N}}\right)^2 + b^2}$ in Fig. 4 as a function of convergence angle. Consistent with Equation 1, which predicts a linear dependence of precision on the semi-convergence angle, we use a linear fit for a vs α. An exponential fit is used for parameter b which represents other sources of systematic error.